\documentclass[twocolumn,showpacs,preprintnumbers,amsmath,amssymb,prl]{revtex4-1}
\usepackage{graphicx} 
\usepackage{dcolumn} 
\usepackage{bm} 
\usepackage{braket} 
\usepackage{multirow} 
\usepackage[usenames,dvipsnames]{color}
\usepackage[
colorlinks=true, citecolor=blue, urlcolor=blue, linkcolor=blue,
setpagesize=false, bookmarks=false
]{hyperref}

\usepackage[english]{babel}
\addto{\captionsenglish}{}

\usepackage[resetlabels]{multibib}
\newcites{SM}{Refs in Supplemental Material}

\begin{document}


\title{Charge stiffness and long-range correlation in the optically induced \\ $\eta$-pairing state of  the one-dimensional Hubbard model}
\author{Tatsuya Kaneko$^{1}$, Seiji Yunoki$^{2,3,4}$, and Andrew J. Millis$^{1,5}$}
\affiliation{
$^1$Department of Physics, Columbia University, New York, New York 10027, USA\\
$^2$Computational Condensed Matter Physics Laboratory, RIKEN Cluster for Pioneering Research (CPR), Wako, Saitama 351-0198, Japan\\
$^3$Computational Materials Science Research Team, RIKEN Center for Computational Science (R-CCS), Kobe, Hyogo 650-0047, Japan \\
$^4$Computational Quantum Matter Research Team, RIKEN Center for Emergent Matter Science (CEMS), Wako, Saitama 351-0198, Japan\\
$^5$Center for Computational Quantum Physics, Flatiron Institute, New York, New York 10010, USA
}
\date{\today}


\begin{abstract}  
We show that optical excitation of the Mott insulating phase of the one-dimensional Hubbard model can create a state possessing two of the hallmarks of superconductivity: a nonvanishing charge stiffness and long-ranged pairing correlation. 
By employing the exact diagonalization method, we find that the superposition of the $\eta$-pairing eigenstates induced by the optical pump exhibits a nonvanishing charge stiffness and a pairing correlation that decays very slowly with system size in sharp contrast to the behavior of an ensemble of thermally excited eigenstates, which has a vanishing charge stiffness and no long-ranged pairing correlations. 
We show that the charge stiffness is indeed directly associated with the $\eta$-pairing correlation in the Hubbard model. 
Our finding demonstrates that optical pumping can actually lead to superconducting-like properties on the basis of the $\eta$-pairing states. 
\end{abstract}

\maketitle


A fundamental goal of nonequilibrium physics is to use strong light-matter interactions to create new quantum phases~\cite{To06,BAH17,OK19,Is19}. 
Recent experimental observation of possible light-induced  superconductivity~\cite{FTDetal11,HKNetal14,KHNetal14,MCNetal16,CBJetal18} has attracted much attention and stimulated many theoretical studies~\cite{DCLetal15,OCMetal16,SKGetal16,KBRetal16,KWRetal17,IOI17,MTEetal17,MG17}. 
These studies focus mainly on the possibility that a light pulse can change the Hamiltonian from one with a nonsuperconducting state into one with a superconducting phase anticipated in equilibrium.  
In contrast, in this paper, we show that optical excitation of the Mott insulating phase of the one-dimensional (1D) Hubbard model excites the system into  a state  characterized by two of the hallmarks of superconductivity: a nonvanishing charge stiffness $D$ and a pairing correlation $P_{ij} = \braket{ \hat{c}^\dagger_{i,\downarrow}\hat{c}^\dagger_{i,\uparrow} \hat{c}_{j,\uparrow}\hat{c}_{j,\downarrow}}_{i\ne j}$ that decays very slowly with system size. 
The components of this state are present in the spectrum but do not give rise to superconducting properties in thermal equilibrium; in other words optical excitation reveals a hidden pairing state. 

Kaneko {\it et al.}~\cite{KSSetal19} showed previously that optical pumping the Mott insulating phase of the Hubbard model created a state characterized by  a pairing correlation $P_{ij} $, whose Fourier transform exhibited a very strong peak at the wave vector $q=\pi$, indicating that a pair density wave state was created~\cite{KSSetal19}. The pair density wave state was attributed to the preferential creation, by the nonequilibrium drive, of  Yang's $\eta$-paired states~\cite{Ya89}.  Subsequent work has demonstrated that $\eta$-pairing can be induced by other protocols including injection of doublon-hole pairs~\cite{WLGetal19,LGWetal19}  and effect of dissipation~\cite{TBCetal19,PPS20}. 
These $\eta$-pairing states are characterized by the operators $\hat{\eta}^{+} = \sum_{j} (-1)^j \hat{c}^{\dag}_{j,\downarrow}    \hat{c}^{\dag}_{j,\uparrow}$, $\hat{\eta}^{-} = \left(\hat{\eta}^+\right)^\dagger$, and $\hat{\eta}_{z} = \frac{1}{2} \sum_{j} \left( \hat{n}_{j,\uparrow} + \hat{n}_{j,\downarrow} - 1  \right)$, where the operators obey the standard SU(2) commutation relations and the operator $\hat{\eta}^+$ creates in effect a paired state with a staggered pairing amplitude~\cite{Ya89,EFGetal05}.  
Since the Hubbard Hamiltonian commutes with the operator $\hat{\eta}^2=\frac{1}{2} \left( \hat{\eta}^{+} \hat{\eta}^{-} + \hat{\eta}^{-} \hat{\eta}^{+} \right) + \hat{\eta}_z^2$, Hubbard eigenstates are simultaneously eigenstates of $\hat{\eta}^2$, and  Yang has shown that a Hubbard eigenstate with a nonzero value of $\braket{\hat{\eta}^2}$ has long-ranged pairing correlations $\left<\eta^+_{i}\eta^-_j\right>_{i\ne j}= (-1)^{i+j}\braket{  \hat{c}^\dagger_{i,\downarrow}\hat{c}^\dagger_{i,\uparrow} \hat{c}_{j,\uparrow}\hat{c}_{j,\downarrow} }$~\cite{Ya89}.  

While  previous work reveals that the  pump electric field  induces  $\eta$-pairing states~\cite{KSSetal19},  actual superconducting properties were not established. In this paper, we  employ an eigenstate analysis and systematic finite-size scaling to show  that the photoinduced $\eta$-pairing state  has nonzero charge stiffness and long-ranged pairing correlations, in contrast, for example, to any thermodynamic ensemble average over states at half-filling, which would  yield an ensemble with no charge stiffness~\cite{CGS13,KKM14,JSHetal15,CNP18}. We also determine the optimal pump profile for the $\eta$-pairing and clarify its system size dependence.


We here study the 1D Hubbard model with the nearest neighbor hopping $t_h$ and on-site interaction $U>0$:
\begin{align}
\hat {\mathcal{H}} = - t_h \sum_{j=1}^{L}\sum_{\sigma}  \left( {\hat c}_{j,\sigma}^{\dag}{\hat c}_{j+1,\sigma} + {\rm H.c.} \right)
 + U \sum_{j=1}^{L}  {\hat n}_{j, \uparrow}  {\hat n}_{j, \downarrow},
\label{eq_Hub}
\end{align}
where ${\hat c}_{j,\sigma}$ (${\hat c}_{j,\sigma}^{\dag}$) is the annihilation (creation) operator for an electron at site $j$  with spin $\sigma$~($= \uparrow, \downarrow$) and ${\hat n}_{j, \sigma}={\hat c}_{j,\sigma}^{\dag}{\hat c}_{j,\sigma}$.
We specialize to the half-filled case with the number of electrons in each spin channel, $N_{\sigma}=L/2$ (number of sites $L$ is taken to be even).
Since $[ \hat{\cal{H}}, \hat{\eta}^2] = [\hat{\cal{H}} , \hat{\eta}_z] =0$, any eigenstate of $\hat{\cal{H}}$ is also the eigenstate $\ket{\eta, \eta_z}$ of ${\hat{\eta}}^2$ and $\hat{\eta}_z$ with the eigenvalues $\eta(\eta+1)$ and $\eta_z$, respectively.
At half-filling, the allowed eigenvalues  $\ket{\eta, \eta_z}$ are  $\eta=0,1,2,\cdots, {L}/{2}$ and $\eta_z=0$.

A time-dependent external field $A(t)$ is introduced via the Peierls substitution $t_h {\hat c}_{j,\sigma}^{\dag}{\hat c}_{j+1,\sigma} \rightarrow t_h e^{iA(t)}{\hat c}_{j,\sigma}^{\dag}{\hat c}_{j+1,\sigma}$.
We use a pump pulse given as $A(t) = A_0 e^{-(t-t_0)^2/(2\sigma_p^2)} \cos \left[ \omega_p (t-t_0)  \right]$ with amplitude $A_0$, frequency $\omega_p$, and pulse width $\sigma_p$ centered at time  $t_0$ ($>0$)~\cite{ceha}.
We assume that for $t=0$ the system is in the Mott insulating ground state and evolve the state forward in time using $\mathcal{\hat{H}}(t)$, which is $\hat{\mathcal{H}}$ with the time-dependent hopping.
We employ the time-dependent exact diagonalization (ED) method~\cite{PL86,MA06} for a finite-size cluster with periodic boundary conditions (PBC) and the state at time $t$ is indicated by $\ket{\Psi(t)}$.
For $t-t_0 \gg\sigma_p$, the resulting state is projected onto the eigenstates $\ket{\psi_m}$ (eigenenergies $\varepsilon_m$) of the unperturbed Hubbard model, obtained by full (exact) diagonalization.
For each eigenstate, we directly calculate the $\eta$-pairing eigenvalue $\eta(\eta+1)$.
We compute the charge stiffness~\cite{CS} for each eigenstate $\ket{\psi_m}$ from
\begin{align}
D_m=\frac{L}{2}\left. \frac{\partial^2\varepsilon_m(\Phi)}{\partial \Phi^2} \right|_{\Phi=0}, 
\end{align} 
with twisted boundary conditions (TBC), where the phase $\Phi$ is introduced via a vector potential 
$A_{\rm twist}=\Phi/L$~\cite{Ko59,Castella95}.    
Details of the method and TBC are given in the Supplemental Material \cite{SM}.


\begin{figure}[b]
\begin{center}
\includegraphics[width=\columnwidth]{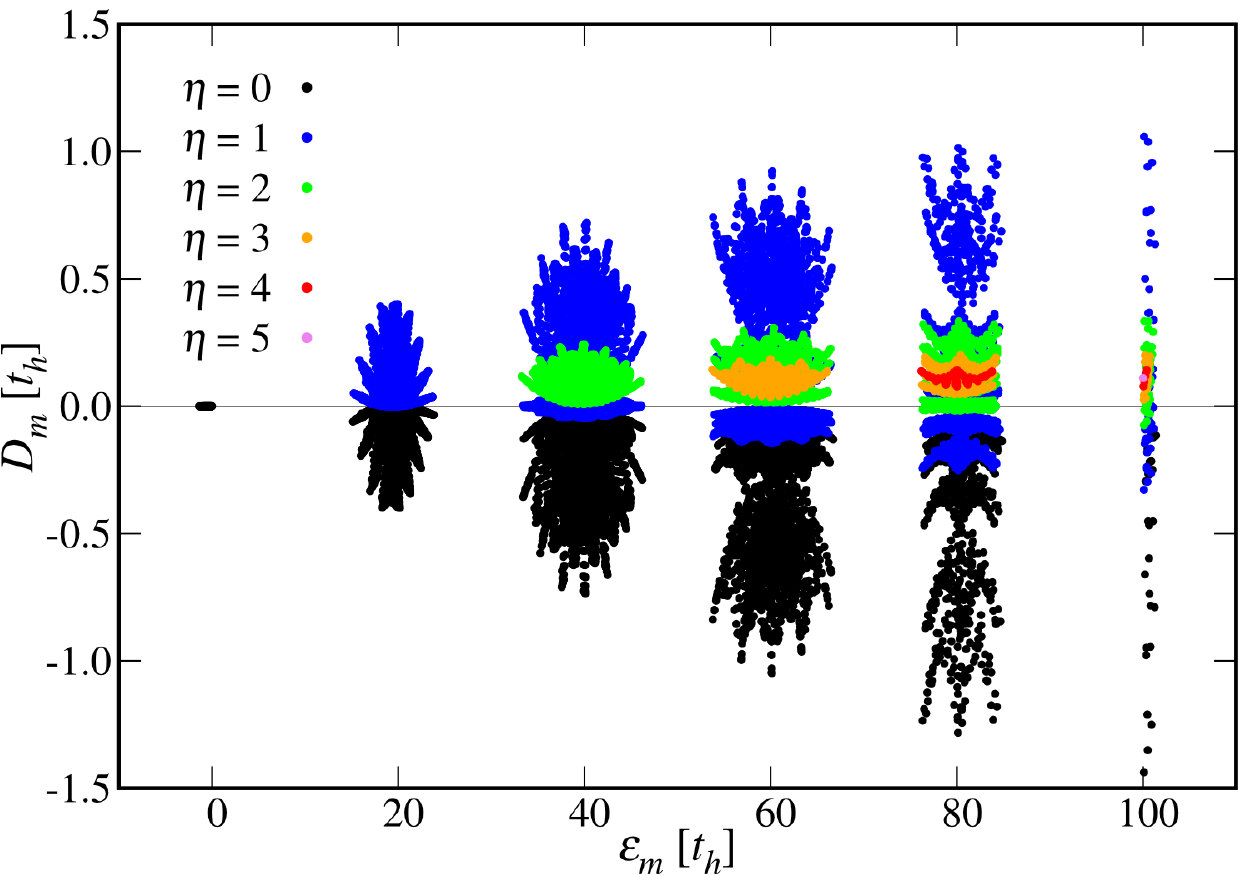}
\caption{
Charge stiffness $D_m$ of the eigenstates $\ket{\psi_m}$ (eigenenergies $\varepsilon_m$) in the half-filled Hubbard chain calculated by the ED method for $L=10$ ($N_{\uparrow}=N_{\downarrow}=5$) at $U=20t_h$.
The colors of the points indicate the values of $\eta$.
}
\label{fig:Drude_eigen}
\end{center}
\end{figure}

Figure~\ref{fig:Drude_eigen} shows the calculated stiffnesses $D_m$ for all eigenstates in the half-filled Hubbard chain at a large value of the interaction $U$.
The eigenstates are grouped into sectors corresponding to different  numbers of doubly occupied sites.
Significantly, most of $D_m$ for the $\eta$-pairing eigenstates ($\eta>0$) are positive, but most of $D_m$ for the non-$\eta$-pairing eigenstates ($\eta=0$) are negative.
The sum of $D_m$ over all eigenstates is zero, because $S(\eta) = \sum_m D_m(\eta)$, the sum of the charge stiffness of all eigenstates with the same $\eta$, satisfies $S(\eta=0) + \sum_{\eta=1}^{L/2} S(\eta) = 0$~\cite{SM}.
This implies that the thermal ensemble at infinite temperature cannot have perfect conducting behavior.
We find numerically that the sum of $D_m$ over all eigenstates in a given double occupancy sector is also zero, and the sum of $D_m$ over all eigenstates within a given small energy range is close to zero.
This strongly suggests that the thermal average of charge stiffness is zero in equilibrium at any temperature as theoretically expected~\cite{CGS13,KKM14,JSHetal15,CNP18}.
To obtain $D>0$ in the half-filled Hubbard chain, one must prepare an ensemble in which $\eta$-pairing ($\eta>0$) eigenstates have larger weight than $\eta=0$ eigenstates.
We next show that photoexcitation produces just such an ensemble.


\begin{figure}[t]
\begin{center}
\includegraphics[width=\columnwidth]{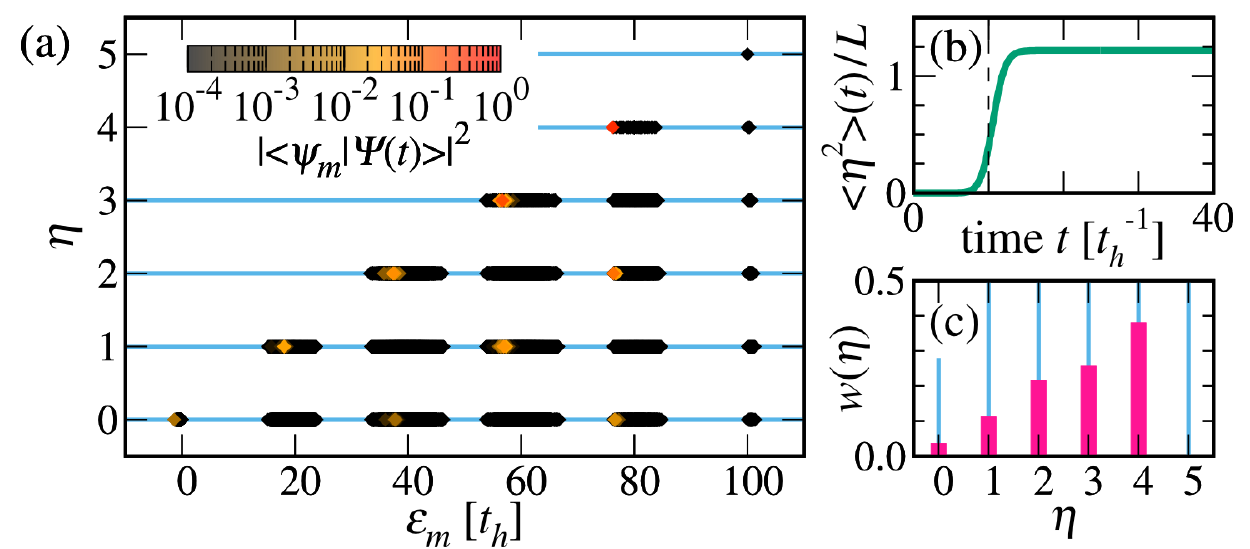}
\\
\includegraphics[width=\columnwidth]{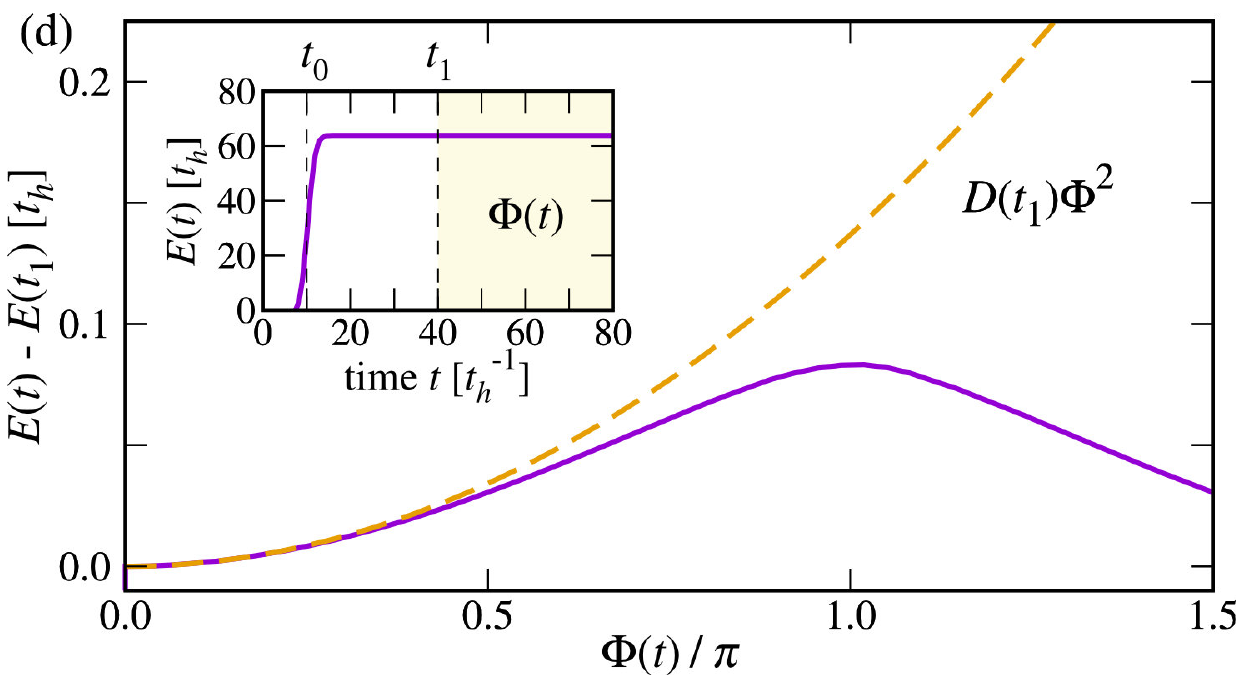}
\caption{
(a)~All eigenenergies $\varepsilon_m$ and eigenvalues $\eta$ for the eigenstates $\ket{\psi_m}$ of the half-filled Hubbard Hamiltonian $\hat{\cal{H}}$ at $U=20t_h$ and $L=10$ ($N_{\uparrow}=N_{\downarrow}=5$) with PBC.
The color of each point indicates the weight $|\braket{\psi_m|\Psi(t)}|^2$ of the eigenstate $\ket{\psi_m}$ in the photoinduced state $\ket{\Psi(t)}$ at $t=40/t_h$ for $A(t)$ with $A_0=0.3$, $\omega_p=19.36t_h$, $\sigma_p=2/t_h$, and  $t_0=10/t_h$. 
(b) Time evolution of $\braket{\hat{\eta}^2}(t)/L=\braket{\Psi(t)|\hat{\eta}^2|\Psi(t)}/L$ 
for the same model parameters in (a). 
(c) Total weight $w(\eta)$ of $|\braket{\psi_m|\Psi(t)}|^2$ over the states $\ket{\psi_m}$ with the same number $\eta$ in (a). Note that $\sum_{\eta=0}^{L/2}w(\eta)=1$.
(d) Time evolution of the energy $E(t) = \braket{\Psi(t)|\hat{\mathcal{H}}(t)|\Psi(t)}$ under the time-dependent flux $\Phi(t)=\theta(t-t_1)\times[ \delta \Phi \cdot  (t-t_1) ]$ applied after the pulse irradiation ($t_1 > t_0$).
The dashed line indicates $\Delta E(\Phi) = D(t_1) \Phi^2/L$ with the charge stiffness $D(t)$ at $t=t_1$ evaluated by $D(t) = \sum_{m} D_m |\braket{\psi_m|\Psi(t)}|^2 $ (see the text).
The inset shows $E(t)$ in the whole energy scale.
The results are calculated using the ED method with $\delta \Phi =0.5\times 10^{-3}$ 
and $t_1 = 40/t_h$ in $\Phi(t)$. 
}
\label{fig:D-photoinduced}
\end{center}
\end{figure}

Before showing $D(t)$, we review the photoinduced state $\ket{\Psi(t)}$ and its weight distribution.
As shown in Fig.~\ref{fig:D-photoinduced}(b), the external pulse $A(t)$ induces an $\eta$-pairing correlation $\braket{\hat{\eta}^2}(t) = \braket{\Psi(t)| \hat{\eta}^2 | \Psi(t)} = \braket{\Psi(t)| \hat{\eta}^{+}\hat{\eta}^{-} | \Psi(t)}$ at half-filling, corresponding to the enhancement of the superconducting correlation at momentum $q=\pi$ shown in Ref.~\cite{KSSetal19}.
Figure.~\ref{fig:D-photoinduced}(a) shows the weight distribution of the eigenstates $\ket{\psi_m}$ in the photoinduced state $\ket{\Psi(t)}$, where the color of each point indicates the weight $ |\braket{\psi_m|\Psi(t)}|^2$ and the total weight is shown as a function of 
$\eta$ in Fig.~\ref{fig:D-photoinduced}(c).
These results clearly show that photoexcitation preferentially induces eigenstates $\ket{\psi_m}$ with $\eta>0$,  explaining the large value of $\braket{\hat{\eta}^2}(t)$ observed in the photoinduced state $\ket{\Psi(t)}$.
This photoinduced nonthermal distribution implies $D(t)>0$.

To verify the stiffness $D(t)>0$ in this photoinduced state $\ket{\Psi(t)}$,  we apply the time-dependent 
flux $A(t)=\Phi(t)/L$, given by $\Phi(t) = \theta(t-t_1)\times[ \delta \Phi \cdot  (t-t_1) ]$, beginning at time $t_1$ long after the pump pulse ($t_1-t_0 \gg \sigma_p$),  where $\theta(t)$ is the Heaviside step function and $\Phi(t)$ increases linearly in time with slope $\delta \Phi$, corresponding to an electric field $\frac{\partial A(t)}{\partial t}\propto \delta \Phi$.
To estimate the stiffness in the photoinduced state $\ket{\Psi(t)}$, we compute the energy 
$E(t) = \braket{\Psi(t)|\hat{\mathcal{H}}(t)|\Psi(t)}$ under the time-dependent flux $\Phi(t)$.
As shown in Fig.~\ref{fig:D-photoinduced}(d), the curvature of $E(t)$ is positive with respect to $\Phi(t)$, indicating $D(t)>0$.
To identify the curvature of the energy $E(t)$ at $\Phi=0$, we should notice that the charge stiffness $D(t) = \sum_{m} |c_m(t)|^2 D_m$ can also be evaluated directly from the weight $|c_m(t)|^2 = |\braket{\psi_m|\Psi(t)}|^2$ in the photoinduced state $\ket{\Psi(t)}$.
Comparing with $\Delta E(\Phi)= D(t_1)\Phi^2/L$, the energy curve $E(t)$ at $\Phi(t) \sim 0$ is perfectly fitted by the stiffness $D(t_1)$ evaluated from the photoinduced weight distribution.
Therefore, the photoinduced state $\ket{\Psi(t)}$ has a stiffness $D(t)>0$.


The above results demonstrate an association between a nonthermal distribution of states with $\braket{\hat{\eta}^2}\ne 0$ and a nonvanishing charge stiffness.
We now show that these two factors are also associated with long-ranged $\eta$-pairing correlation.
First, we see this association in Yang's maximally $\eta$-paired state $\ket{\phi_{N_{\eta}}} \propto \left( \hat{\eta}^{+} \right)^{N_{\eta}} \ket{0}$ generated from the vacuum $\ket{0}$~\cite{Ya89}.
For this state, Yang showed that the $\eta$-pairing correlation is distance independent and of infinite range with $\braket{\phi_{N_{\eta}}| \hat{\eta}^+_i \hat{\eta}^-_j  |\phi_{N_{\eta}}}_{i \ne j}=\frac{N_{\eta}\left( L - N_{\eta}\right)}{L\left( L-1\right)}$~\cite{Ya89}.
Here we find that the charge stiffness $D_{\eta}$ for Yang's $\eta$-pairing state $\ket{\phi_{N_{\eta}}}$ satisfies $D_{\eta}=   4 J_{\rm ex} \braket{\phi_{N_{\eta}}| \hat{\eta}_i^{+} \hat{\eta}_j^{-}  |\phi_{N_{\eta}}}_{i\ne j} > 0$ 
with the exchange interaction $J_{\rm ex} = 2t_h^2 / U$ (see details in the Supplemental Material~\cite{SM}), which directly associates the charge stiffness with the long-ranged pairing correlation.  

Our numerical evidence strongly suggests that this association is valid beyond Yang's $\eta$-pairing state.
To discuss this, let us review the ingredients of $\braket{\hat{\eta}^2}$.
At half-filling ($\eta_z=0$), the algebra of $\eta$ operators implies
\begin{align}
\braket{\hat{\eta}^2} = L n_d +  \sum_{i \ne j}  \braket{ \hat{\eta}^+_i \hat{\eta}^-_j  }
\label{eq:eta_nd-LR}
\end{align}
with the double occupancy $n_d = \frac{1}{L}\sum_{j} \braket{ \hat{n}_{j,\uparrow}\hat{n}_{j,\downarrow}}$.
From the analysis of the eigenstates, we can show $\braket{\hat{\eta}^2}(n_d) \equiv \frac{1}{\mathcal{N}_{n_d}} \sum_{m} \braket{\psi_m | \hat{\eta}^2 | \psi_m}_{n_d}= L n_d$ in each double occupancy ($n_d$) sector, where $\mathcal{N}_{n_d}$ is the number of the eigenstates and the suffix $n_d$ indicates the eigenstate within the $n_d$ sector (see the Supplemental Material~\cite{SM}).
We can also show that the average of $\braket{\psi_m | \hat{\eta}^2 | \psi_m}$ over all Hubbard eigenstates at half-filling is $\braket{\hat{\eta}^2}_{\rm avr.}/L=0.25$, which is same with the double occupancy $n_d=0.25$ at infinite temperature.
Comparing with Eq.~(\ref{eq:eta_nd-LR}), these relations strongly suggest that a thermal distribution of the eigenstates has no long-range $\eta$-pairing correlation.
However, we find for the optically generated state $\ket{\Psi(t)}$ that  $\braket{\hat{\eta}^2} > L n_d$ 
[see, e.g., Fig.~\ref{fig:D-photoinduced}(b), where $\braket{\hat{\eta}^2}(t)/L > 1$], which implies contributions from nonlocal pairing correlations $\braket{ \hat{\eta}^+_i \hat{\eta}^-_j }_{i\ne j}$ in Eq.~(\ref{eq:eta_nd-LR}).

\begin{figure}[t]
\begin{center}
\includegraphics[width=\columnwidth]{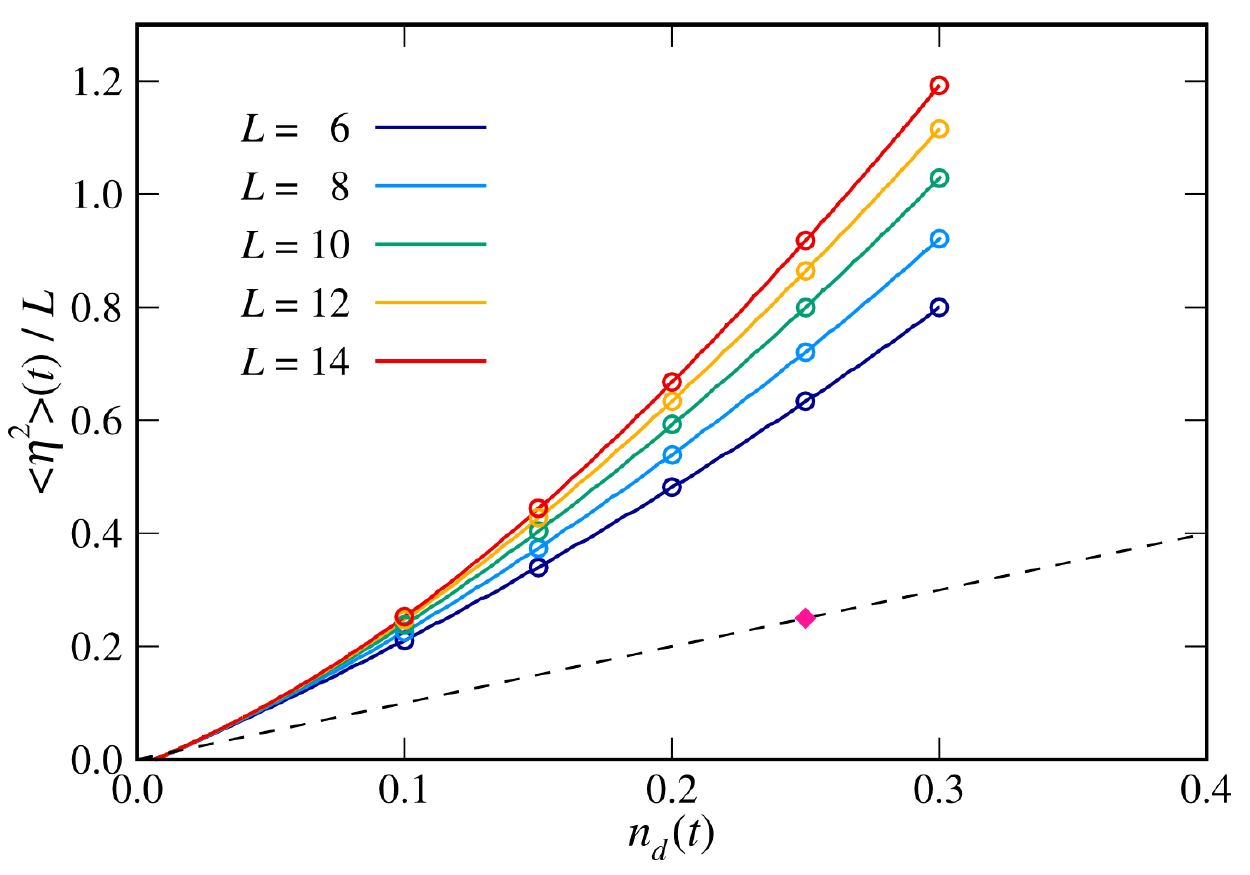}
\caption{
$\braket{\hat{\eta}^2}(t)/L$ as the function of $n_d(t)$ in the half-filled Hubbard chain at $U=20t_h$  with $\omega_p/t_h=18.68$, 19.11, 19.36, 19.54, and 19.66 for $L=6$, 8, 10, 12, and 14, respectively.
The dashed line is $\braket{\hat{\eta}^2}(n_d)/L= n_d$.
The diamond indicates $\braket{\hat{\eta}^2}/L= n_d=0.25$, which is the average of $\braket{\psi_m | \hat{\eta}^2 | \psi_m}$ over all Hubbard eigenstates at half-filling ($N_{\uparrow}=N_{\downarrow}=L/2$).
The results are calculated by the ED method under PBC with $A_0=0.3$, $\sigma_p=2/t_h$, and  $t_0=10/t_h$ in $A(t)$.
}
\label{fig:PI-Ldep_eta-nd}
\end{center}
\end{figure}

We now analyze the spatial correlations and finite size effects in the photoinduced state. 
One trivial finite size effect is a weak size dependence of the optimal photoexcitation frequency $\omega_p$.
For each system size, we calculate $\braket{\hat{\eta}^2}(t)$ with different $\omega_p$  (see the Supplemental Material \cite{SM}).
Here we present results obtained at the optimal $\omega_p$ for each size.
We represent the amount of optical excitation by the induced double occupancy in Fig.~\ref{fig:PI-Ldep_eta-nd}, by plotting $\braket{\hat{\eta}^2}(t)/L$ as a function of $n_d(t)=  \frac{1}{L}\sum_{j} \braket{\Psi(t)| \hat{n}_{j,\uparrow}\hat{n}_{j,\downarrow} | \Psi(t)}$.
Note that here we consider a fixed pump strength $A_0$, which produces time-dependent $n_d(t)$ and $\braket{\hat{\eta}^2}(t)$.  Equivalent results could be obtained by $A_0$ dependence considering the long-time limits of $n_d(t)$ and $\braket{\hat{\eta^2}}(t)$  (see the Supplemental Material \cite{SM}).
Figure~\ref{fig:PI-Ldep_eta-nd} reveals two important results: $\braket{\hat{\eta}^2}(t)/L$ under photoexcitation is systematically greater than $n_d$ (dashed line), indicating that the $\braket{ \hat{\eta}^+_i \hat{\eta}^-_j  }_{i \ne j}$ term in Eq.~(\ref{eq:eta_nd-LR}) is nonzero, and the difference from $n_d$ increases with increasing system size $L$.
Examination of Eq.~(\ref{eq:eta_nd-LR}) indicates that this increase must correspond to the development of long-range correlation.
In comparison with an average with thermal distribution of the eigenstates, where $\braket{\hat{\eta}^2}/ L \sim n_d$ and $D\sim 0$ are expected, $\braket{\hat{\eta}^2}(t)/ L > n_d(t)$  in Fig.~\ref{fig:PI-Ldep_eta-nd} implies a nonthermal distribution induced by optically preferential $\eta$-pairing states, 
which gives rise to a nonvanishing charge stiffness $D(t)>0$ (see, e.g., Fig.~\ref{fig:D-photoinduced}).

\begin{figure}[t]
\begin{center}
\includegraphics[width=\columnwidth]{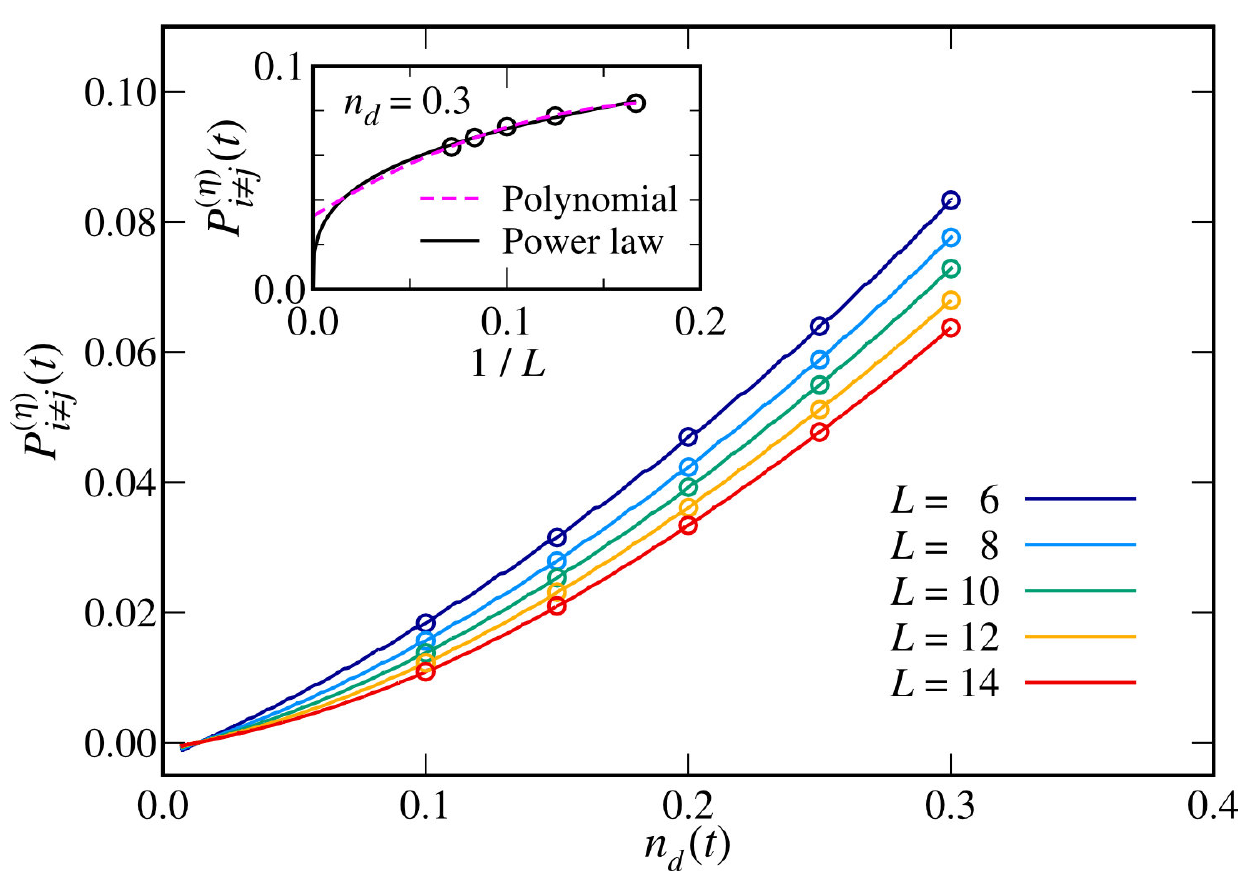}
\caption{
Time-dependent $\eta$-pairing correlation $P_{i\ne j}^{(\eta)}(t)$ as the function of the double occupancy $n_d(t)$ in the half-filled Hubbard chain at $U=20t_h$ with $\omega_p/t_h=18.68$, 19.11, 19.36, 19.54, and 19.66 for $L=6$, 8, 10, 12, and 14, respectively.
Inset: Size dependence of $P_{i\ne j}^{(\eta)}(t)$ at $n_d (t)= 0.3$.
The dashed and solid lines are polynomial and power-law fittings, respectively.
The results are calculated by the ED method under PBC with $A_0=0.3$, $\sigma_p=2/t_h$, and  $t_0=10/t_h$ in $A(t)$.
}
\label{fig:eta-photoinduced}
\end{center}
\end{figure}

To further understand the pairing correlation, we define the quantity
\begin{align}
P_{i\ne j}^{(\eta)}(t) = \frac{1}{L^2} \sum_{i \ne j}  \braket{\Psi(t)| \hat{\eta}^+_i \hat{\eta}^-_j | \Psi(t)}.
\label{eq:eta_LR}
\end{align} 
When long-ranged $\eta$-pairing correlation is formed, 
$P_{i\ne j}^{(\eta)}(t)$ remains nonzero with increasing system size $L$, corresponding to $\braket{\hat{\eta}^2} \sim \sum_{i \ne j}  \braket{ \hat{\eta}^+_i \hat{\eta}^-_j } \propto L^2$.
For Yang's $\eta$-pairing state $\ket{\phi_{\eta}}$, $P_{i\ne j}^{(\eta)} = 0.25$ at $n_d=0.5$  regardless of the system sizes.
In Fig.~\ref{fig:eta-photoinduced}, we show $P_{i\ne j}^{(\eta)}(t)$ with the different system size $L$.
We see for the optically created state that the magnitude is $P_{i\ne j}^{(\eta)}(t)\sim 0.07$ at $n_d(t)=0.3$, which is comparable to the value in Yang's maximally $\eta$-paired state.
The value of $P_{i\ne j}^{(\eta)}(t)$ varies slowly with system size.
The inset of Fig.~\ref{fig:eta-photoinduced} shows the $L$ dependence of $P_{i\ne j}^{(\eta)}(t)$ at $n_d(t)=0.3$.
While the range of system sizes accessible to us is too small to make a definitive statement, 
the results are consistent with either a nonzero extrapolation to the $L\rightarrow \infty$ limit or $P_{i\ne j}^{(\eta)} \propto L^{-\alpha}$ with $\alpha\sim 0.3$ corresponding to very slowly decaying power-law pairing correlation (quasi-long-range order).

\begin{figure}[t]
\begin{center}
\includegraphics[width=\columnwidth]{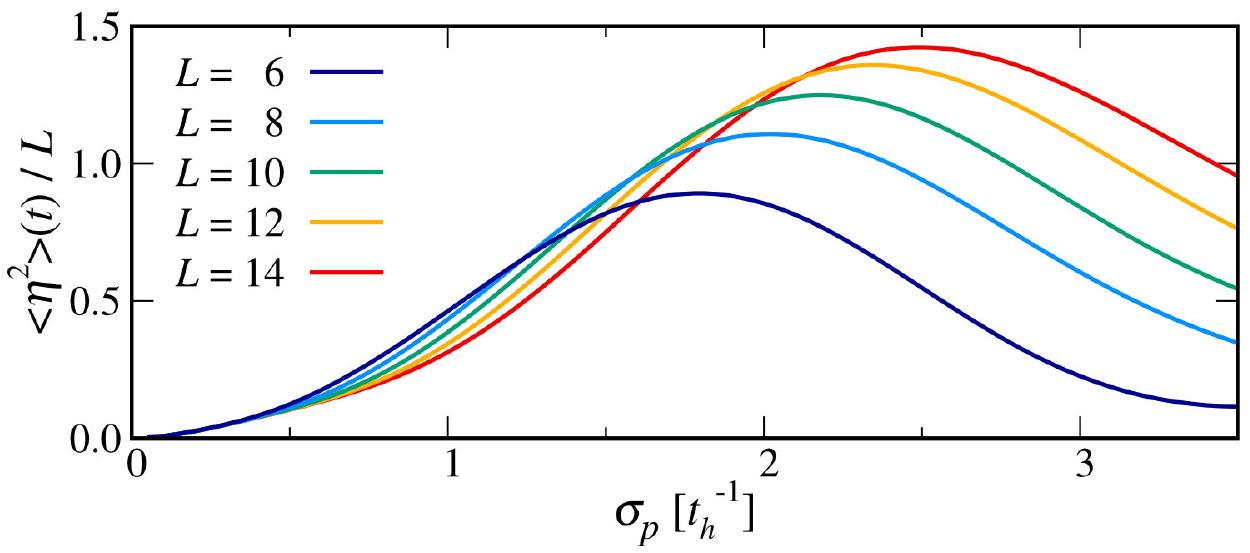}
\caption{
Dependence of $\braket{\hat{\eta}^2}(t)/L$ on pump width $\sigma_p$, computed  for  the half-filled Hubbard chain using the ED method with  PBC and $U=20t_h$ at $t = 10\sigma_p + 10/t_h$ after the pump maximum. The pump frequencies  are $\omega_p/t_h=18.68$, $19.11$, $19.36$, $19.54$, and $19.66$ for $L=6$, $8$, $10$, $12$, and $14$, respectively, and the other pump parameters are $A_0=0.3$ and  $t_0=5\sigma_p$. 
}
\label{fig:eta_sigma_p}
\end{center}
\end{figure}

Finally, we comment on the pulse width $\sigma_p$ dependence. Since a high-temperature ensemble is expected in the limit of $\sigma_p \rightarrow \infty$ at $\omega_p \sim U$~\cite{HMEetal18}, there must be an optimal value of $\sigma_p$ for the enhancement of the $\eta$-pairing correlation.  Figure~\ref{fig:eta_sigma_p} shows the $\sigma_p$ dependence of $\braket{\hat{\eta}^2}(t)/L$ computed at a long time  after the optical pump.  The optimal pump width and the maximal value of $\braket{\hat{\eta}^2}(t)/L$ increase with increasing $L$, which is consistent with the idea that the pump produces a state with long-ranged correlations.


In conclusion, we have shown from finite system numerics along with a scaling analysis of the system size dependence that optical excitation of the 1D Hubbard model creates a state possessing two of the hallmarks of superconductivity: a nonvanishing charge stiffness and long-ranged pairing correlation. The fundamental reason is that optical excitation preferentially creates $\eta$-pairing states, which as we have shown here via an eigenstate analysis have a positive stiffness with typical values of $D \sim J_{\rm ex} = 2t_h^2/U$. This work extends the previous study~\cite{KSSetal19} showing that optical excitation can  induce  $\eta$-pairing correlations by demonstrating  that the nonequilibrium ensemble created by the drive in fact has superconducting properties. 

While the 1D Hubbard model we used here is in several respects a highly simplified description of real materials,  it can be realized in cold atomic gasses and our results provide predictions for experiments in these systems. But, more fundamentally,  we believe that our results are important because they provide an existence proof  that nonequilibrium drive can create a state with superconducting properties in an originally nonsuperconducting system. Our work provides new understanding of the qualitative properties of light-induced superconductivity, and may serve as a base for future research  including both a more detailed examination of the properties of the light-induced superconducting state and extensions  to higher dimensions and richer models~\cite{SMY20}.


The authors thank S. Ejima, D. Gole\v{z}, and T. Shirakawa for fruitful discussion.
This was supported in part by Grants-in-Aid for Scientific Research from JSPS 
(Projects No.~JP18K13509, No.~JP18H01183, and No.~JP20H01849) of Japan.
A.J.M. was supported by the Basic Energy Sciences program of the U.S. Department of Energy under Grant No.~DE-SC0018218.
T.K. was supported by the JSPS Overseas Research Fellowship.


\bibliography{References}

\begin{thebibliography}{46}
\bibitem [{ceh()}]{ceha}%
  \BibitemOpen
  \href@noop {} {}\bibinfo {note} {The light velocity $c$, the elementary
  charge $e$, the Planck constant $\hbar$, and the lattice constant are set to
  1. The hopping $t_h$ ($t_h^{-1}$) is a unit of energy (time).}\BibitemShut
  {Stop}%
\bibitem [{CS()}]{CS}%
  \BibitemOpen
  \href@noop {} {}\bibinfo {note} {The charge stiffness $D$ is the coefficient
  of a delta-function term in the optical response: ${\rm Re} \left[
  \sigma(\omega)\right] = 2\pi D\delta(\omega) + \sigma_{\rm reg}(\omega)$, and
  $D>0$ implies dissipationless charge transport.}\BibitemShut {Stop}%
\end{thebibliography}

\clearpage

\appendix
\renewcommand\thesection{}
\renewcommand{\theequation}{S\arabic{equation}}
\setcounter{equation}{0}
\renewcommand\thefigure{S.\arabic{figure}}
\setcounter{figure}{0}
\renewcommand{\bibnumfmt}[1]{[S#1]}
\renewcommand{\citenumfont}[1]{S#1}

\subsection*{ Supplemental Material}


\subsection{1. Time-dependent exact diagonalization method} 

For the time evolution of the state $\ket{\Psi(t)}$, we employ the time-dependent exact diagonalization (ED) method. 
The detail is described in Ref.~\citeSM{KSSetal19S}. 
Our time-dependent ED method is based on the Lanczos algorithm and the time evolved state with a short time step $\delta t$ is calculated in the corresponding Krylov subspace generated with $M_{\rm L}$ Lanczos iterations~\citeSM{PL86S,MA06S,HI16S}. 
Here, we adopt $\delta t = 0.001 / t_h$ and $M_{\rm L} = 15$ for the time evolution. 
We assume $\ket{\Psi(t=0)}=\ket{\psi_0}$ as the initial state, where $\ket{\psi_0}$ is the ground (Mott insulating) state of the Hubbard Hamiltonian $\hat{\cal{H}}$.


\subsection{2. Twisted boundary conditions}

In order to estimate the charge stiffness, we consider the one-dimensional (1D) Hubbard model with the flux $\Phi$, described by 
\begin{align}
\hat {\mathcal{H}}_{\Phi} = - t_h \sum_{j,\sigma}  \left( e^{i\frac{\Phi}{L}}{\hat c}_{j,\sigma}^{\dag}{\hat c}_{j+1,\sigma} + {\rm H.c.} \right)  
 + U \sum_{j}  {\hat n}_{j, \uparrow}  {\hat n}_{j, \downarrow}.             
\label{PI-eta_eq1}                    
\end{align}
Notice that through a transformation $\hat{\tilde{c}}_{j,\sigma}=e^{i\frac{\Phi}{L}R_j}\hat{c}_{j,\sigma}$, where $R_j$ is the location of site $j$, $\hat {\mathcal{H}}_{\Phi}$ is transformed to $\hat {\mathcal{H}}$ defined in Eq.~(1) 
in the main text with a simple substitution $\hat{c}_{j,\sigma} \to \hat{\tilde{c}}_{j,\sigma}$. However, since the operator $\hat{\tilde{c}}_{j,\sigma}$ satisfies $\hat{\tilde{c}}_{L+1,\sigma}=e^{i\Phi}\hat{\tilde{c}}_{1,\sigma}$, the transformed Hamiltonian $\hat {\mathcal{H}}$ has to satisfy twisted boundary conditions (TBC) with the phase $\Phi$, in stead of periodic boundary conditions (PBC). 

Even in the presence of the flux $\Phi$, we can still define the $\eta$-pairing operators. 
With the local pair operators $\hat{\eta}^+_j=(\hat{\eta}^-_j)^{\dag}=(-1)^j \hat{c}^{\dag}_{j,\downarrow} \hat{c}^{\dag}_{j,\uparrow}$ and $\hat{\eta}^z_j=\frac{1}{2}\left( {\hat n}_{j, \uparrow} + {\hat n}_{j, \downarrow} -1 \right)$, $\eta$-pairing operators under the flux $\Phi$ are given by $\hat{\eta}_{\Phi}^{\pm} = \sum_j e^{\mp i\frac{2\Phi}{L}R_j} \hat{\eta}^\pm_j$ and $\hat{\eta}_z=\sum_j\hat{\eta}^z_j$~\citeSM{ND04S}.   
These operators also satisfy the $SU(2)$ commutation relations $\left[ \hat{\eta}_{z},\hat{\eta}_{\Phi}^{\pm} \right] = \pm \hat{\eta}_{\Phi}^{\pm}$ and $\left[ \hat{\eta}_{\Phi}^{+},\hat{\eta}_{\Phi}^{-} \right] = 2\hat{\eta}_z$. 
However, in contrast to the $\eta$-pairing operators at $\Phi=0$, $[\hat {\mathcal{H}}_{\Phi}, \hat{\eta}_{\Phi}^{+} \hat{\eta}_{\Phi}^{-}]\ne 0$ for arbitrary $\Phi$. 
The Hamiltonian $\hat {\mathcal{H}}_{\Phi}$ commutes with $\hat{\eta}_{\Phi}^2=\frac{1}{2} \left( \hat{\eta}_{\Phi}^{+} \hat{\eta}_{\Phi}^{-} + \hat{\eta}_{\Phi}^{-} \hat{\eta}_{\Phi}^{+} \right) + \hat{\eta}_z^2$  only at $\Phi = n \pi$ ($n = 0, \pm1, \pm2, \cdots$)~\citeSM{ND04S}. 

\begin{figure}[t]
\begin{center}
\includegraphics[width=\columnwidth]{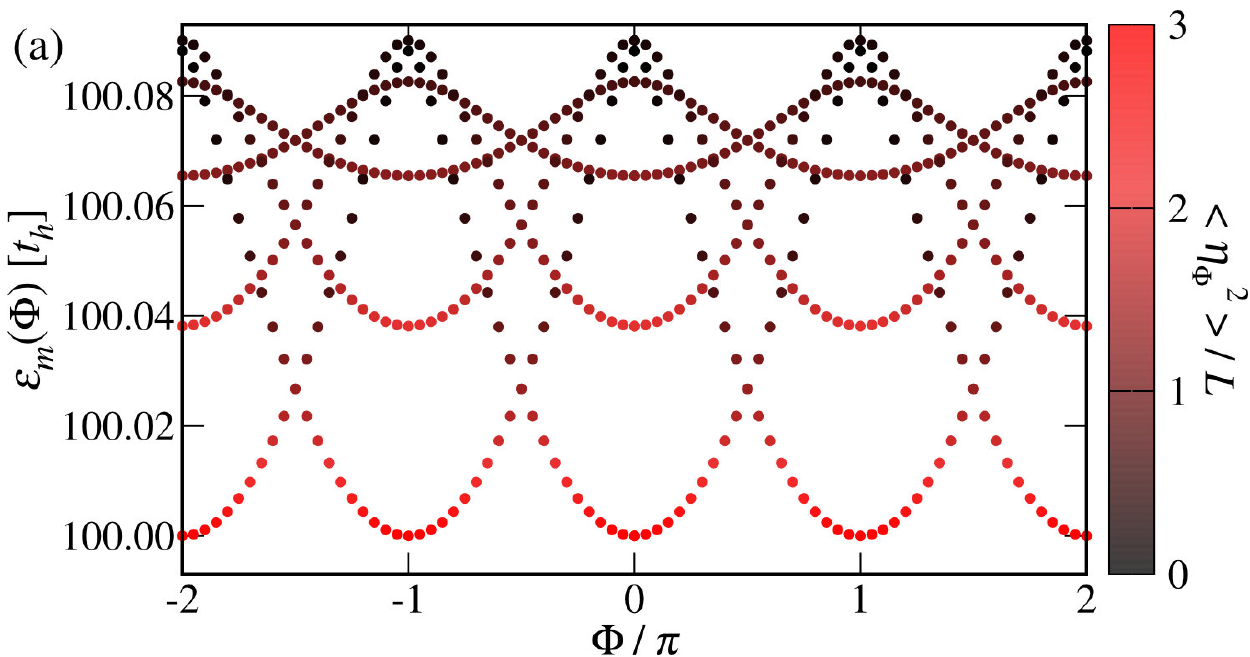}
\\
\includegraphics[width=\columnwidth]{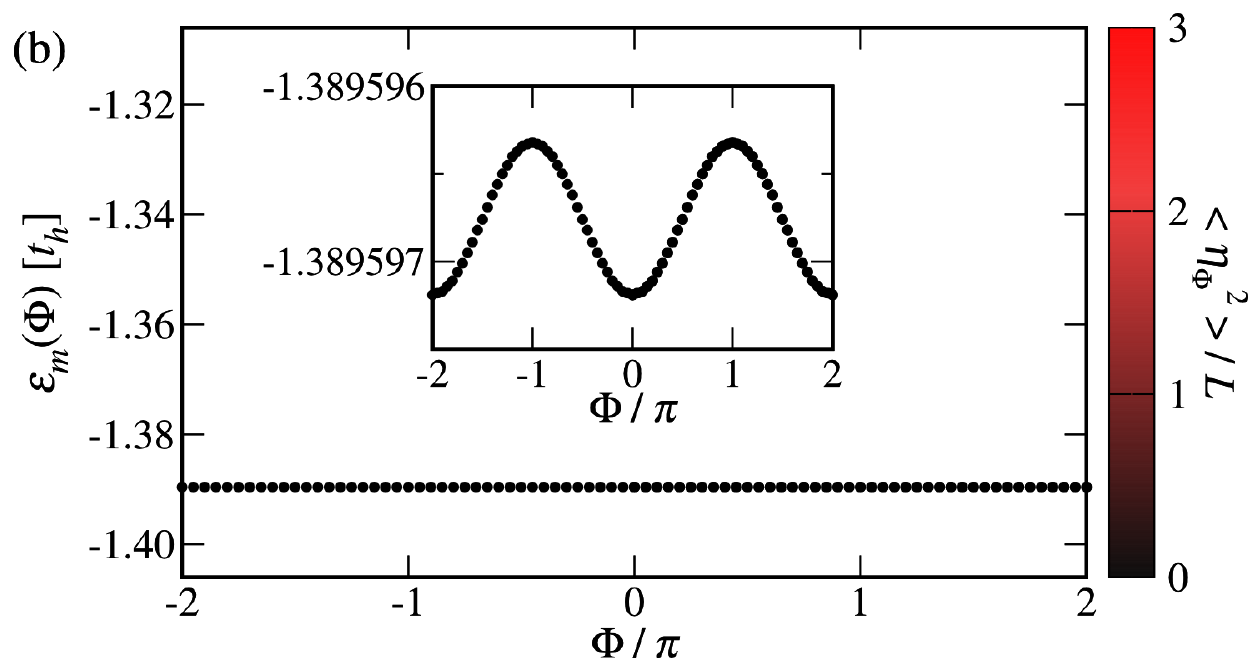}
\caption{
Eigenenergies $\varepsilon_m(\Phi)$ in the half-filled Hubbard model as a function of the flux $\Phi$  calculated by the ED method for $L=10$ at $U=20t_h$. 
(a) $\varepsilon_m(\Phi)$ in the vicinity of Yang's $\eta$-pairing state at $\varepsilon_m(\Phi)=UL/2$ and (b) $\varepsilon_m(\Phi)$ for the ground state are plotted. 
The color of each point indicates the $\eta$-pairing correlation $\braket{\psi_m(\Phi)|\hat{\eta}_{\Phi}^2|\psi_m(\Phi)}/L$ for the eigenstate $\ket{\psi_m(\Phi)}$. 
The inset of (b) is the enlarged plot of (b). 
Notice that the scale of the vertical axis in the inset of (b) is orders of magnitude smaller.
} 
\label{fig:ene_flux}
\end{center}
\end{figure}

In Fig.~\ref{fig:ene_flux}(a), we calculate the eigenenergies $\varepsilon_m(\Phi)$ of the half-filled Hubbard Hamiltonian $\hat {\mathcal{H}}_{\Phi}$, only showing the eigenenergies in the vicinity of Yang's $\eta$-pairing state at $\varepsilon_m(\Phi=0)=UL/2$.  
The color of each point indicates the $\eta$-pairing correlation $\braket{\psi_m(\Phi)| \hat{\eta}_{\Phi}^2 | \psi_m(\Phi)}= \braket{\psi_m(\Phi)| \hat{\eta}_{\Phi}^{+} \hat{\eta}_{\Phi}^{-}  | \psi_m(\Phi)}$ of the eigenstate $\ket{\psi_m (\Phi)}$.  
We can clearly observe that the energy curves $\varepsilon_{m}(\Phi)$ have minima with a period of $\pi$, indicating the signature of the flux quantization with charge $2e$. 
This should be contrasted with the behavior of $\varepsilon_{m}(\Phi)$ for the eigenstates around the Mott insulating ground state, which oscillates with a period of $2\pi$, shown in the inset of Fig.~\ref{fig:ene_flux}(b).

Under TBC, the charge stiffness $D_m$ of the eigenstate $\ket{\psi_m}$ is given by 
\begin{align}
D_{m}
=   \frac{L}{2} \left. \frac{\partial^2 \varepsilon_m(\Phi)}{\partial \Phi^2}\right|_{\Phi=0}. 
\end{align}
Because $\varepsilon_m(\Phi) = \braket{\psi_{m}(\Phi) | \hat{\mathcal{H}}_{\Phi} | \psi_{m}(\Phi)}$, the second order perturbation analysis with respect to $\Phi$ provides
\begin{align}
D_{m}
=  - \frac{1}{2L} \braket{\psi_{m} | \hat{T} | \psi_{m}}
- \frac{1}{L}   
\sum_{n (\ne m)}  
\frac{ |\braket{\psi_{n}  | \hat{J} | \psi_{m} } |^2}{\varepsilon_n-\varepsilon_m}, 
\label{eq:D-eigen}
\end{align}
where the kinetic operator $\hat{T}$ and current operator $\hat{J}$ are defined as
\begin{align}
&\hat{T}
= -  t_h \sum_{j,\sigma}  \left( \hat{c}_{j,\sigma}^{\dag} \hat{c}_{j+1,\sigma} +  \hat{c}_{j+1,\sigma}^{\dag}\hat{c}_{j,\sigma} \right) ,
\\
&\hat{J}
= - it_h \sum_{j,\sigma}  \left( \hat{c}_{j,\sigma}^{\dag}\hat{c}_{j+1,\sigma}  - \hat{c}_{j+1,\sigma}^{\dag}\hat{c}_{j,\sigma}  \right).                  
\end{align}
In this paper, the charge stiffness $D_m$ is computed by $D_m=(L/2)\times [\varepsilon_m(\delta \Phi)+\varepsilon_m(-\delta \Phi)-2\varepsilon_m(0)]/(\delta\Phi)^2$ with $\delta \Phi=0.001$.


\subsection{3. Charge stiffness of Yang's $\eta$-pairing state}

In this section, we derive the charge stiffness of Yang's maximally $\eta$-paired state
\begin{align}
\ket{\phi_{N_{\eta}}} = \frac{1}{\sqrt{\mathcal{C}_{N_{\eta}}}}\left( \hat{\eta}^{+} \right)^{N_{\eta}} \ket{0}, 
\end{align}
where $N_{\eta}$ is the number of $\eta$-pairs and  $\mathcal{C}_{N_{\eta}}= N_{\eta} ! \prod^{N_{\eta}}_{k=1} (L-k+1)$. 
$\ket{\phi_{N_{\eta}}}$ is an eigenstate of the Hubbard model with the eigenenergy $N_{\eta} U$~\citeSM{Ya89S}. 
Since $\braket{\phi_{N_{\eta}} | \hat{T} | \phi_{N_{\eta}}} = 0$ in Eq.~(\ref{eq:D-eigen}), the stiffness $D_{\eta}$ is given by
\begin{align}
D_{\eta} =  - \frac{1}{L} 
\sum_{n}  \frac{ |\braket{\psi_n | \hat{J} | \phi_{N_{\eta}}} |^2}{\varepsilon_n - N_{\eta} U}. 
\end{align}
Because Yang's $\eta$-pairing state satisfies $\hat{\mathcal{H}}\hat{J}\ket{\phi_{N_{\eta}}}=(N_{\eta}-1)U\hat{J}\ket{\phi_{N_{\eta}}}$, the normalized state 
\begin{align}
\ket{\tilde{\phi}_{N_{\eta}-1}}
= \frac{1}{\sqrt{\braket{\phi_{N_{\eta}}|\hat{J}^2|\phi_{N_{\eta}} }}}\hat{J}\ket{\phi_{N_{\eta}}}
\end{align}
is also the eigenstate of the Hubbard Hamiltonian $\hat{\mathcal{H}}$ with the eigenenergy $(N_{\eta}-1)U$. 
Therefore, $D_{\eta}$ becomes
\begin{align}
D_{\eta}
=   \frac{1}{LU}\braket{\phi_{N_{\eta}}|\hat{J}^2|\phi_{N_{\eta}}}.
\label{eq:D-eta-J^2}
\end{align}
By using the commutation relations between $\hat{\eta}^{\pm}$ and $\hat{J}$ previously derived in Ref.~\citeSM{KSSetal19S}, we obtain  
\begin{align}
\braket{\phi_{N_{\eta}}|\hat{J}^2|\phi_{N_{\eta}}}
= 8t_h^2 \frac{N_{\eta}\left( L - N_{\eta}\right)}{ L-1} . 
\label{eq:J^2}
\end{align}
Finally, combining Eqs.~(\ref{eq:D-eta-J^2}) and (\ref{eq:J^2}), we obtain
\begin{align} 
D_{\eta} = 4J_{\rm ex} \frac{N_{\eta}\left( L - N_{\eta}\right)}{ L(L-1)} ,
\end{align}
where we introduced the exchange interaction $J_{\rm ex} = 2t_h^2/U$. 
The stiffness $D_{\eta}$ of Yang's $\eta$-pairing state is thus characterized by the exchange interaction $ J_{\rm ex}$ and the number of $\eta$-pairs $N_{\eta}$. 
At half-filling ($N_{\eta}=L/2$), $D_{\eta}=   J_{\rm ex} L/(L-1)$, which becomes $D_{\eta} =  J_{\rm ex}$ in the $L\rightarrow \infty$ limit. 

Importantly, since the off-diagonal long-range order in Yang's $\eta$-pairing state is characterized by~\citeSM{Ya89S}
\begin{align}
\braket{\phi_{N_{\eta}}| \hat{\eta}^+_i \hat{\eta}^-_j  |\phi_{N_{\eta}}}_{i \ne j}= \frac{N_{\eta}\left( L - N_{\eta}\right)}{L\left( L-1\right)}, 
\end{align}
the charge stiffness $D_{\eta}$ becomes
\begin{align}
D_{\eta}
=   4 J_{\rm ex} \braket{\phi_{N_{\eta}}| \hat{\eta}_i^{+} \hat{\eta}_j^{-}  |\phi_{N_{\eta}}}_{i\ne j} .
\label{eq:suppl_Deta}
\end{align}
Therefore, the $\eta$-pairing correlation is directly associated with the stiffness $D_{\eta}$ and Yang's state $\ket{\phi_{N_{\eta}}}$ has the nonvanishing stiffness $D_{\eta}>0$.


\subsection{4. Charge stiffness and superfluid density}

While the charge stiffness $D$ is well defined even in one dimension, the evaluation of the superfluid density (and thus Meissner effect) requires a system in more than one spatial dimension~\citeSM{SWZ93S,Resta18S}. 
Although the long-range $\eta$-pairing correlation is a necessary and sufficient condition for superconductivity, 
$D\ne0$ is a sufficient condition because a perfect metal can also exhibit $D\ne0$~\citeSM{SWZ93S,Resta18S}. 
However, this is excluded in Yang's $\eta$-pairing state because the charge stiffness in Eq.~(\ref{eq:suppl_Deta}) is directly associated with the long-range $\eta$-pairing correlation. 
Furthermore, the corresponding energy curve $\varepsilon_m(\Phi)$ has minima with a period of $\pi$ as shown in Fig.~\ref{fig:ene_flux}(a), revealing the signature of the flux quantization with charge $2e$, not with charge $e$ expected for a perfect metal~\citeSM{Byers61S}. 
The direct calculation of superfluid density in the photoexcited state in a higher dimensional system is an interesting extension and this is left for a future study.


\subsection{5. Sum rule for charge stiffness}

Here, we show the following relation
\begin{align}
S = \sum_{m}D_{m} = 0. 
\end{align}
In this and the next sections, to indicate the $\eta$ degrees of freedom explicitly, we describe the eigenstate as $\ket{\psi_{m}; \eta,\eta_z}$. 
From Eq.~(\ref{eq:D-eigen}), the charge stiffness of the eigenstate $\ket{\psi_{m}; \eta,\eta_z}$ at half-filling ($\eta_z=0$) is given by 
\begin{align}
D_{m}(\eta)
= &- \frac{1}{2L}  \braket{\psi_{m};\eta,0 | \hat{T} | \psi_{m};\eta,0 }
\notag \\
&-  \frac{1}{L}  \sum_{n}  
\frac{ |\braket{\psi_{n}; \eta',0  | \hat{J} | \psi_{m}; \eta,0 } |^2}{\varepsilon_n(\eta')-\varepsilon_m(\eta)}.
\end{align}
Because of the selection rule $\braket{\psi_{n}; \eta',0  | \hat{J} | \psi_{m}; \eta,0 } = 0$ when $\eta' \ne \eta \pm 1$~\citeSM{KSSetal19S,FKOetal19S}, the charge stiffness is given by
\begin{align}
D_{m}(\eta)  =  
& - \frac{1}{2L}  \braket{\psi_{m}; \eta,0 | \hat{T} | \psi_{m}; \eta,0 }
\notag \\
&- \frac{1}{L}  {\sum_n}' 
\frac{ |\braket{\psi_{n}; \eta-1,0  | \hat{J} | \psi_{m}; \eta,0 } |^2}{\varepsilon_n(\eta-1)-\varepsilon_{m}(\eta)}
\notag \\
&- \frac{1}{L}  {\sum_n}' 
\frac{ |\braket{\psi_{n};\eta+1,0  | \hat{J} | \psi_{m}; \eta,0 } |^2}{\varepsilon_n(\eta+1)-\varepsilon_m(\eta)}. 
\label{eq:D-eigen_wSL}
\end{align} 
To avoid possible confusion, here we indicate by the prime the sum over all energy eigenstates $\ket{\psi_{n}; \eta,\eta_z}$ with a particular value of $\eta$ ($\eta_z=0$ at half-filling). 

Here, we define the sum of the charge stiffness for the eigenstates with the same number of $\eta$ as 
\begin{align}
S(\eta) = {\sum_{m}}' D_{m}(\eta). 
\end{align} 
We divide the contribution from $\hat{T}$ and $\hat{J}$ operators as 
\begin{align}
S(\eta) = S_{T}(\eta) + S_{J}(\eta)
\end{align} 
and discuss  $S_{T}(\eta)$ and  $S_{J}(\eta)$ separately. 
First, $S_{T}(\eta)$ is given by 
\begin{align}
 S_T (\eta)
 = - \frac{1}{2L}  {\sum_{m}}' \braket{\psi_{m}; \eta,0 | \hat{T} | \psi_{m}; \eta,0 }. 
\end{align} 
Because we assume the particle-hole symmetric structure with $\hat{T} = -2t_h \sum_{k,\sigma} \cos (k) c^{\dag}_{k,\sigma} c_{k,\sigma}$, the sum for all eigenstates at half-filling satisfies 
\begin{align}
\sum_{\eta=0}^{L/2} {\sum_{m}}' \braket{\psi_{m}; \eta,0 | \hat{T} | \psi_{m}; \eta,0 } = 0, 
\end{align} 
and we thus obtain 
\begin{align}
\sum_{\eta=0}^{L/2} S_T (\eta)
= 0. 
\label{eq:sum_ST}
\end{align} 

Next, $S_{J}(\eta)$ is given by 
\begin{align}
S_J(\eta)
& = \frac{1}{L}  {\sum_{m,n}}' 
\frac{ |\braket{\psi_{n}; \eta,0  | \hat{J} | \psi_{m}; \eta-1,0 } |^2}{\varepsilon_{n}(\eta)-\varepsilon_m(\eta-1)}
\notag \\
&-  \frac{1}{L}  {\sum_{m,n}}' 
\frac{ |\braket{\psi_{n}; \eta+1,0  | \hat{J} | \psi_{m}; \eta,0 } |^2}{\varepsilon_n(\eta+1)-\varepsilon_{m}(\eta)}, 
\end{align} 
where the sum $S_J(\eta)$ is characterized by the transitions $\eta-1 \rightarrow \eta$ and $\eta \rightarrow \eta + 1$. 
Introducing the function 
\begin{align}
F(\eta+1,\eta) = 
 \frac{1}{L}  {\sum_{m,n}}' 
 \frac{ |\braket{\psi_{n}; \eta+1,0  | \hat{J} | \psi_{m}; \eta,0 } |^2}{\varepsilon_n(\eta+1)-\varepsilon_m(\eta)}, 
\end{align}
the sum $S_J(\eta)$ is given by
\begin{align}
S_J(\eta) =   F(\eta,\eta-1)-F(\eta+1,\eta) , 
\end{align}
where $S(L/2) =   F(L/2,L/2-1)$ and $S(0) =  -F(1,0)$. 
Because $F(\eta+1,\eta)$ in $S_J(\eta)$ and $S_J(\eta+1)$ cancels each other, the sum of $S_J(\eta)$ for all $\eta$ becomes
\begin{align}
\sum_{\eta = 0}^{L/2} S_J(\eta) = \sum_{\eta = 0}^{L/2} \left[ F(\eta,\eta-1) -F(\eta+1,\eta)  \right] = 0. 
\label{eq:sum_SJ}
\end{align}
Combining Eqs.~(\ref{eq:sum_ST}) and (\ref{eq:sum_SJ}), we finally obtain 
\begin{align}
S = \sum_{\eta = 0}^{L/2} S(\eta) 
= \sum_{\eta = 0}^{L/2} \left[ S_T(\eta)  +  S_J(\eta) \right]
 = 0. 
\end{align} 
Therefore, the sum of $D_m$ over all eigenstate is zero. 
Because of the derivation shown above, $S=0$ also gives us the following interesting relation: 
\begin{align}
\sum_{\eta = 1}^{L/2} S(\eta) =  -S(0). 
\end{align}
The sum of the charge stiffness for the $\eta=0$ eigenstates and that for the $\eta>0$ eigenstates have the opposite sign. 
The calculated $D_m$ shown in the main text satisfies this relation, and most of $D_m$ for the $\eta>0$ eigenstates are positive and most of  $D_m$ for the $\eta=0$ eigenstates are negative.


\subsection{6. Average of $\braket{\hat{\eta}^2}$}

\begin{figure}[t]
\begin{center}
\includegraphics[width=\columnwidth]{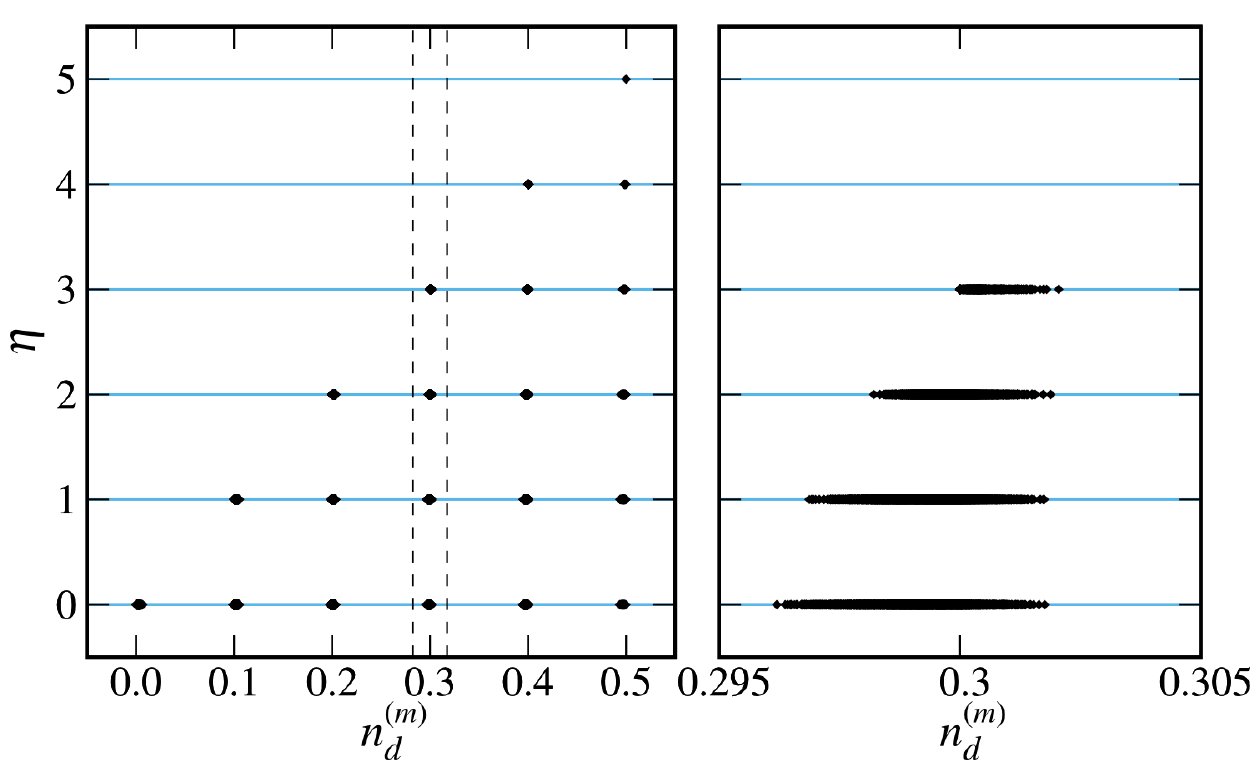}
\caption{
Double occupancies $n_d^{(m)}$ and eigenvalues $\eta$ for the eigenstates $\ket{\psi_m}$ of the half-filled Hubbard model ($N_{\uparrow}=N_{\downarrow}=L/2$) at $U=20t_h$ and $L=10$ under PBC. 
Right panel: $n_d^{(m)}$ and $\eta$ around $n_d=0.3$. 
}
\label{fig:eigen_nd-eta}
\end{center}
\end{figure}

\begin{table}[t]
\caption{Number of the eigenstates for the half-filled Hubbard model with $N_{\uparrow}=N_{\downarrow}=L/2 = 5$. }
\begin{center}
\begin{tabular}{   c  r | r  r  r  r  r  r | r }
 \multicolumn{2}{ c |}{ }& \multicolumn{6}{ c |}{$N_d$ ($= L n_d$)} &  \multirow{2}{*}{} 
  \\  
\makebox[13pt][r]{} &  &  \makebox[26pt][r]{0} & \makebox[26pt][r]{1} & \makebox[26pt][r]{2} & \makebox[26pt][r]{3} & \makebox[26pt][r]{4} & \makebox[26pt][r]{5} & \makebox[28pt]{}
 \\ \hline
 \multirow{6}{*}{\makebox[5pt]{$\eta$}} &5 &  &  &   &  &  & 1 & 1
 \\ 
&4 &  &  &  &  & 90 & 9 & 99
 \\ 
&3 &  &  &   & 1260 & 630 & 35 & 1925
 \\ 
&2 &  &  & 4200 & 6300 & 1800 & 75 & 12375 
 \\ 
&1 &  & 3150 & 12600 & 11340 & 2520 & 90 & 29700
 \\ 
&0 & 252 & 3150 & 8400 & 6300 & 1260 & 42 & 19404  
 \\ \hline
 \multicolumn{2}{ c |}{ } & 252 & 6300 & 25200 & 25200 & 6300 & 252 & 63504 \\
\end{tabular}
\label{table:eigen_nd-eta}
\end{center}
\end{table}

In this section, we estimate the average of $\braket{\psi_m| \hat{\eta}^2 | \psi_m}$ within each double occupancy ($n_d$) sector. 
In Fig.~\ref{fig:eigen_nd-eta}, we show the value of $\eta$ and double occupancy 
\begin{align}
n^{(m)}_d = \frac{1}{L} \sum_{j} \braket{\psi_m| \hat{n}_{j,\uparrow} \hat{n}_{j,\downarrow} | \psi_m}
\end{align}
for all eigenstates $\ket{\psi_m}$ of the ten-site Hubbard ring at half-filling. 
The large $U$ means  the eigenstates are grouped also into sectors of different double occupancies $n_d=N_d/L$ ($N_d = 0, 1, \ldots ,L/2$) together with different values of $\eta$.
The average of $\braket{\psi_m| \hat{\eta}^2 | \psi_m}$ belonging to the $n_d$ sector may be defined as 
\begin{align}
\braket{\hat{\eta}^2} (n_d) 
&= \frac{1}{\mathcal{N}_{n_d}} \sum_{m} \braket{\psi_m| \hat{\eta}^2 | \psi_m}_{n_d} 
\notag \\
&= \frac{1}{\mathcal{N}_{n_d}} 
{ \sum_{\eta=0}^{N_d} {\sum_{\substack{m \\ | n^{(m)}_d \!- n_d | < \Delta n_d}}}^{\hspace{-20pt}\prime} } 
\! \! \! \hspace{20pt} \braket{\psi_m; \eta,0 | \hat{\eta}^2|\psi_m; \eta,0} , 
\end{align}
where we sum up $\braket{\psi_m| \hat{\eta}^2 | \psi_m}$ for the eigenstates  within a range $| n^{(m)}_d \!- n_d | < \Delta n_d$ and $\mathcal{N}_{n_d}$ is the number of the eigenstates within the range. 
Note that, since we intend to estimate the average of $\braket{\psi_m| \hat{\eta}^2 | \psi_m}$ in each $n_d$ sector, we assume an appropriate $\Delta n_d$ (as small as $0.01$ in this case) to pick up all eigenstate belonging to the $n_d$ sector.  
Because $\braket{\psi_m; \eta,0 | \hat{\eta}^2|\psi_m; \eta,0} = \eta(\eta+1)$, we have 
\begin{align}
\braket{\hat{\eta}^2} (n_d) 
= \frac{\displaystyle \sum_{\eta=0}^{N_d} \eta(\eta+1) \mathcal{N}_{\eta,N_d}}{\displaystyle \sum_{\eta=0}^{N_d} \mathcal{N}_{\eta,N_d}}, 
\label{eq:avr_eta}
\end{align}
where $\mathcal{N}_{\eta,N_d}$ is the number of the eigenstates with $\eta$ in the $n_d$ $(=N_d/L)$ sector and $\mathcal{N}_{n_d}=\sum_{\eta}\mathcal{N}_{\eta,N_d}$. 

In Table~\ref{table:eigen_nd-eta}, we show $\mathcal{N}_{\eta,N_d}$, corresponding to Fig.~\ref{fig:eigen_nd-eta}, for the ten-site Hubbard ring at half-filling ($N_{\uparrow}=N_{\downarrow}=L/2$). 
We can show that $\mathcal{N}_{\eta,N_d}$ in Table~\ref{table:eigen_nd-eta} is given as
\begin{align}
\mathcal{N}_{\eta,N_d}
= &\binom{L}{L- 2N_d} \binom{L - 2N_d}{L/2 - N_d} 
\notag \\
&\times \left[ \binom{2N_d}{N_d - \eta} - \binom{2N_d}{N_d - \eta - 1} \right], 
\label{eq:num_eta-nd}
\end{align}
where $\binom{L}{L- 2N_d} \binom{L - 2N_d}{L/2 - N_d}$ is the number of the states on the singly occupied sites and $\binom{2N_d}{N_d - \eta} - \binom{2N_d}{N_d - \eta - 1}$ is the number of the states on the doubly and no occupied sites with the different $\eta$~\citeSM{Ta71S,EKS91S,EKS92S}. 
Combining Eqs.~(\ref{eq:avr_eta}) and (\ref{eq:num_eta-nd}), we obtain 
\begin{align}
\braket{\hat{\eta}^2} (n_d)  = N_d = L n_d.
\label{eq:eta=nd}
\end{align}

We can also show that the average of $\braket{\psi_m| \hat{\eta}^2 | \psi_m}$ over all eigenstates is $\braket{\hat{\eta}^2}_{\rm avr.}/L=0.25$, which is the same as the double occupancy $n_d=0.25$ at infinite temperature.


\subsection{7. Photoinduced $\eta$-pairing}

\begin{figure}[t]
\begin{center}
\includegraphics[width=0.98\columnwidth]{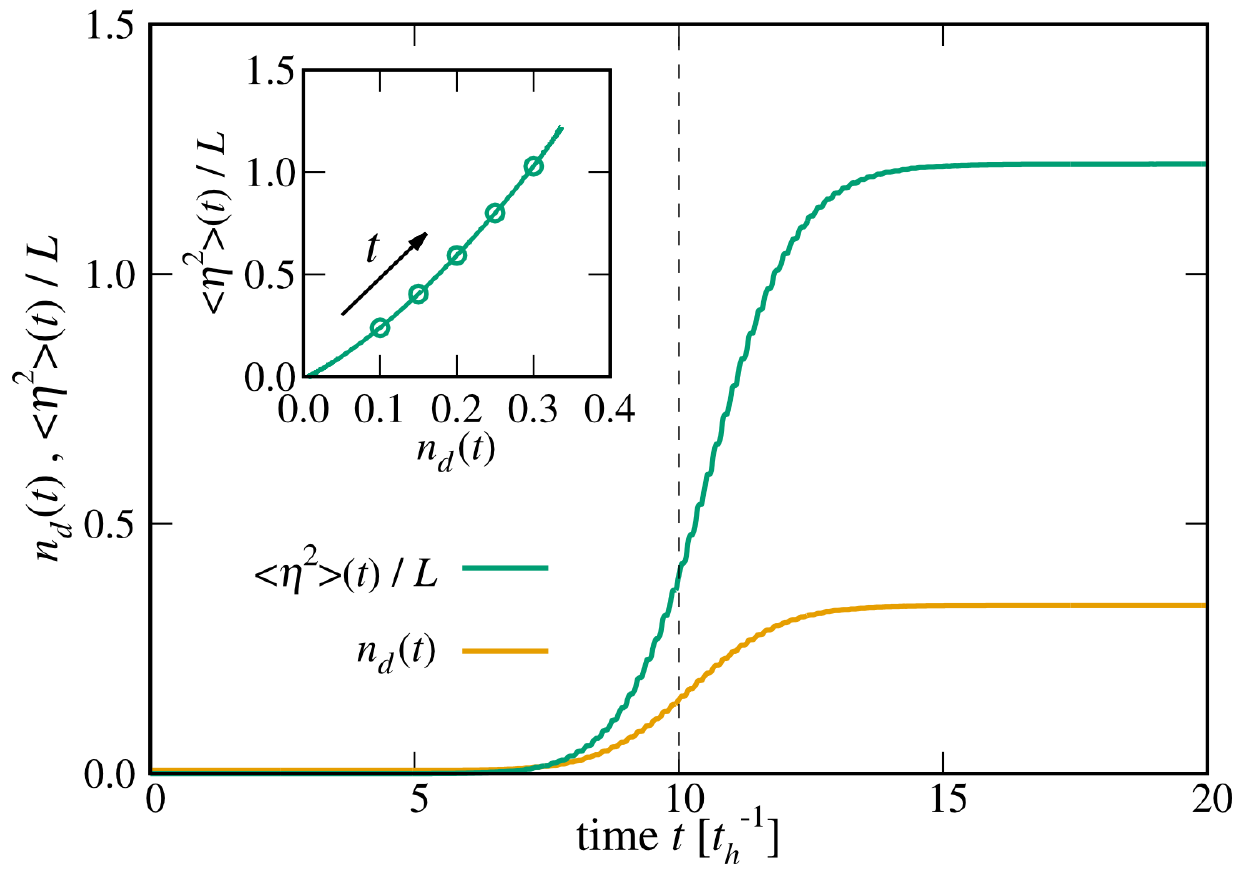}
\caption{
Time evolution of the double occupancy $n_d(t)$ and the $\eta$-pairing correlation $\braket{\hat{\eta}^2}(t)/L$. 
Inset: $\braket{\hat{\eta}^2}(t)/L$ as the function of $n_d(t)$. The arrow indicates the time-evolved direction. 
The results are calculated by the ED method for $L=10$ under PBC at $U=20t_h$ with $A_0=0.3$, $\omega_p=19.36t_h$, $\sigma_p=2/t_h$, and  $t_0=10/t_h$ in $A(t)$.  
}
\label{fig:PI-TE_eta-nd}
\end{center}
\end{figure}

\begin{figure}[t]
\begin{center}
\includegraphics[width=\columnwidth]{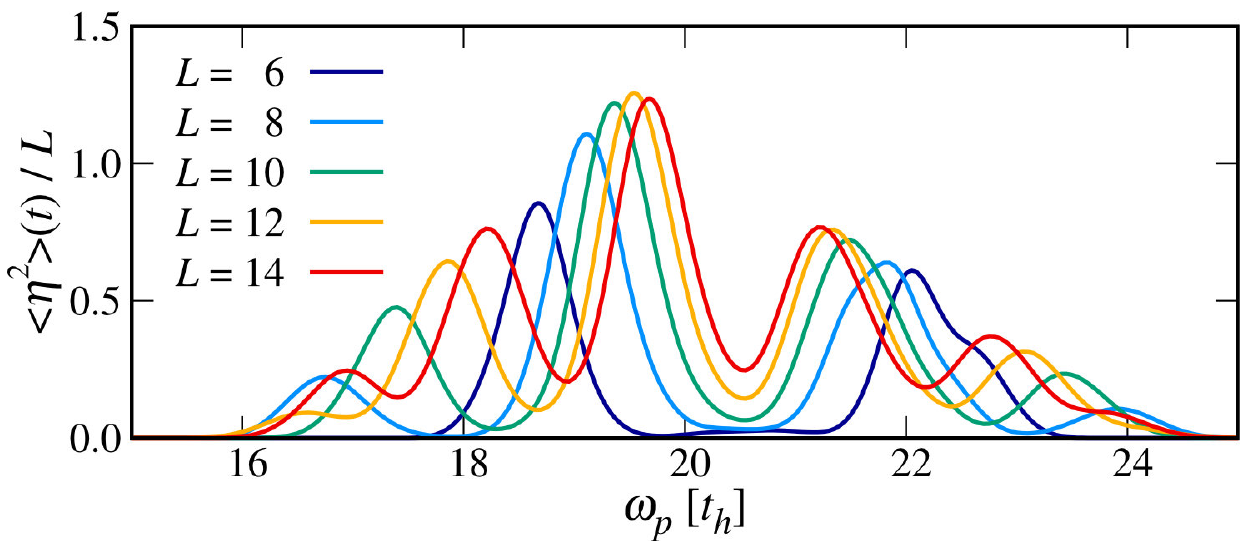}
\caption{
Frequency $\omega_p$ dependence of $\braket{\hat{\eta}^2}(t)/L$ in the half-filled Hubbard chain at $U=20t_h$ after the pulse irradiation ($t=30/t_h$) with different system size $L$.  
The results are calculated by the ED method under PBC with $A_0=0.3$, $\sigma_p=2/t_h$, and  $t_0=10/t_h$ in $A(t)$.  
}
\label{fig:PI-Ldep_eta-wp}
\end{center}
\end{figure}

\begin{figure}[!t]
\begin{center}
\includegraphics[width=0.98\columnwidth]{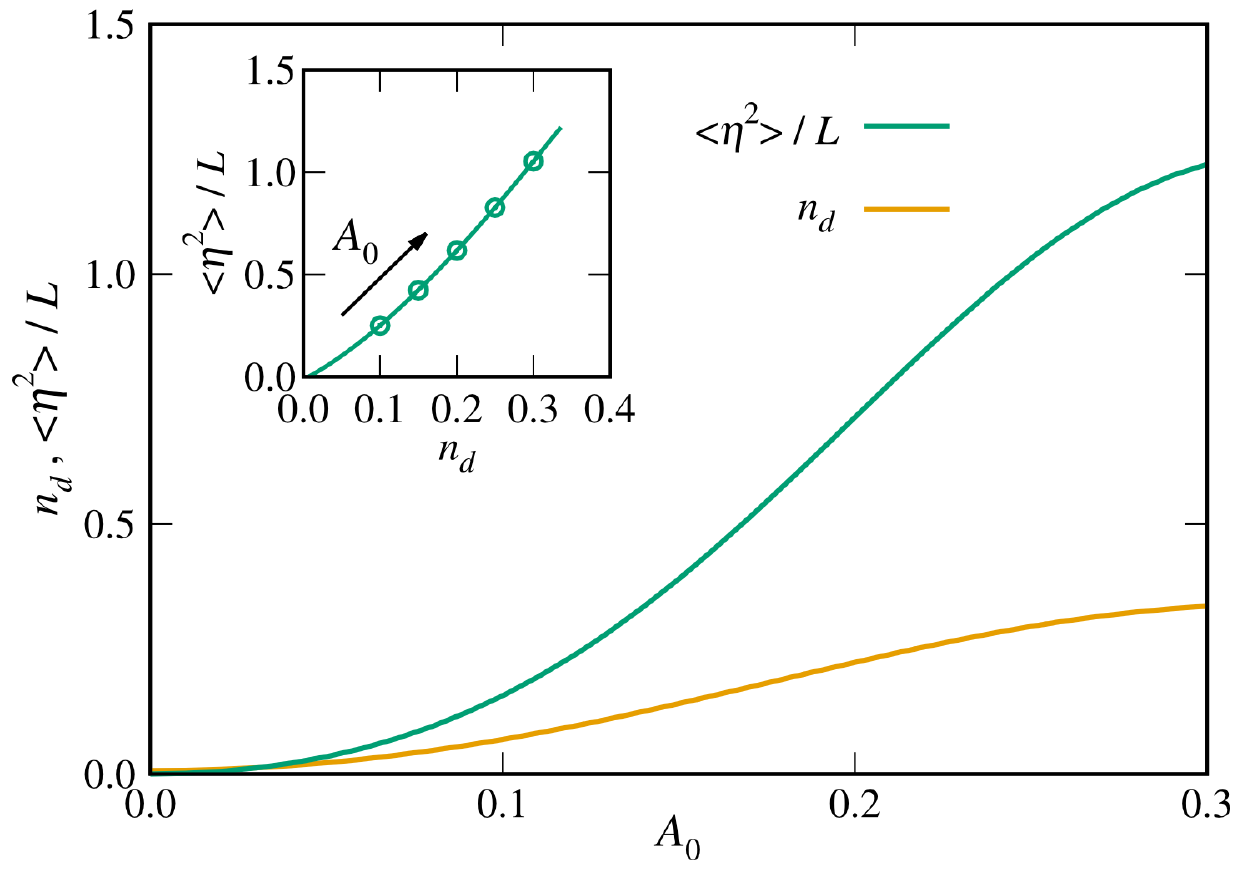}
\caption{
$A_0$ dependence of $n_d$ and $\braket{\hat{\eta}^2}/L$  averaged from $t = 20/t_h$ to $30/t_h$ after pumping. 
Inset: $\braket{\hat{\eta}^2}/L$ as the function of $n_d$. 
Note that these quantities are calculated at the same $A_0$ as shown in the main figure and the arrow indicates the direction of increasing $A_0$. 
The results are calculated by the ED method for $L=10$ under PBC at $U=20t_h$ with $\omega_p=19.36t_h$, $\sigma_p=2/t_h$, and  $t_0=10/t_h$ in $A(t)$.  
}
\label{fig:PI-A_eta-nd}
\end{center}
\end{figure}

\begin{figure}[t]
\begin{center}
\includegraphics[width=0.98\columnwidth]{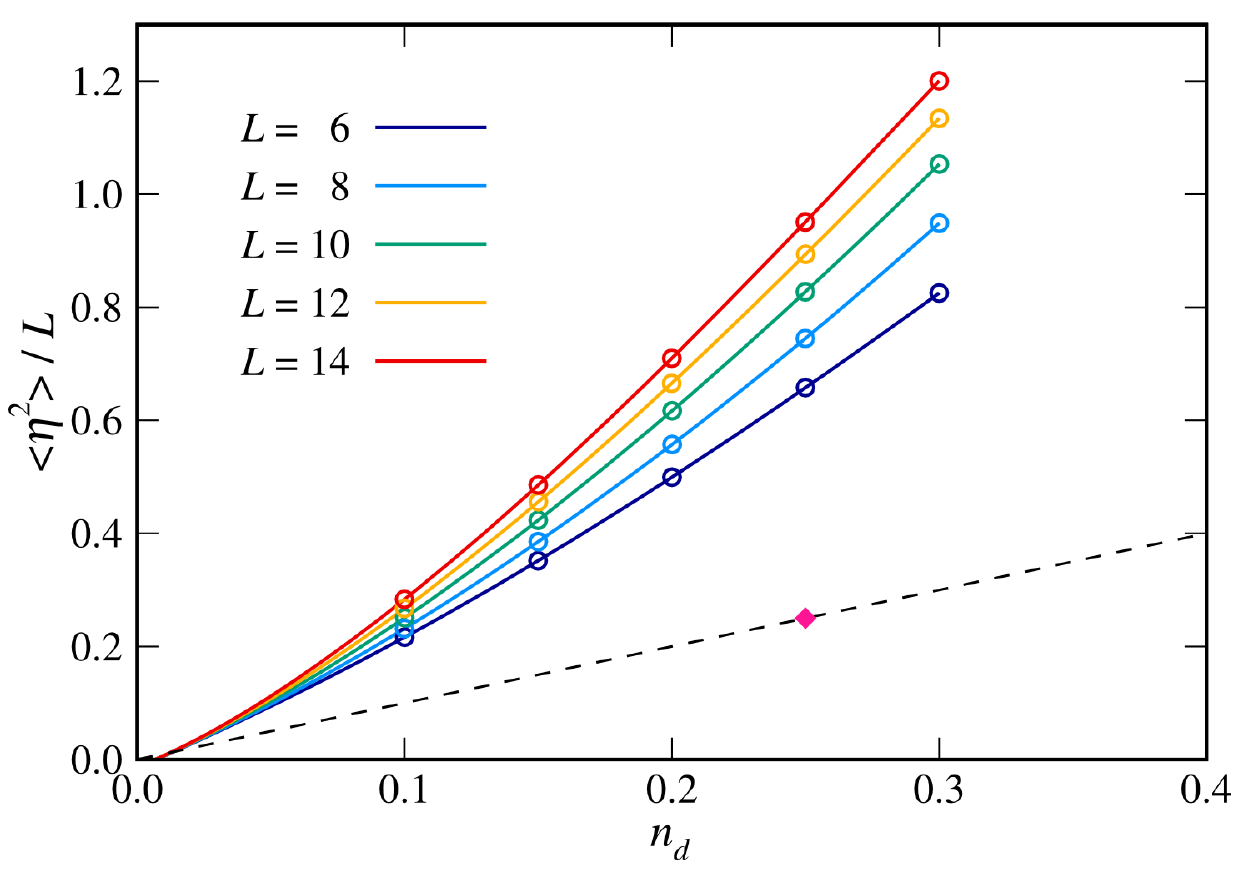}
\caption{
$\braket{\hat{\eta}^2}/L$ as a function of $n_d$ for different values of $A_0$ in the half-filled Hubbard chain at $U=20t_h$. 
Here, these quantities are 
time-averaged from $t = 20/t_h$ to $30/t_h$ after pumping.  
The pulse frequencies $\omega_p/t_h=18.68$, 19.11, 19.36, 19.54, and 19.66 are set for $L=6$, 8, 10, 12, and 14, respectively. 
The dashed line is $\braket{\hat{\eta}^2} \! (n_d) \! / L = n_d$. 
The diamond indicates $\braket{\hat{\eta}^2}/L =  n_d = 0.25$, which is the average of $\braket{\psi_m | \hat{\eta}^2 | \psi_m}$ 
over all Hubbard eigenstates at half-filling ($N_{\uparrow}  =  N_{\downarrow}  =  L/2$). 
The results are calculated by the ED method under PBC with $\sigma_p =  2/t_h$ and  $t_0  = 10/t_h$ in $A(t)$.  
}
\label{fig:PI-Ldep_eta-ndSM}
\end{center}
\end{figure}

Here we provide the supplemental data for the photoinduced $\eta$-pairing state. 
Figure~\ref{fig:PI-TE_eta-nd} shows the time-evolution of the double occupancy
\begin{align}
n_d(t) = \frac{1}{L}\sum_{j} \bra{\Psi(t)} {\hat n}_{j,\uparrow} {\hat n}_{j,\downarrow} \ket{\Psi(t)}
\end{align}
and the $\eta$-pairing correlation 
\begin{align}
\braket{\hat{\eta}^2}(t) = \braket{\Psi(t)| \hat{\eta}^2 | \Psi(t)}. 
\end{align} 
Note that $\braket{\hat{\eta}^2}(t) =  \braket{\Psi (t)| \hat{\eta}^{+} \hat{\eta}^{-}  | \Psi(t)}$  
because $\eta_z = 0$ at half-filling and therefore $2\braket{\hat{\eta}^2}(t)/L$ is exactly the same quantity as the superconducting structure factor $P(q,t)$ at $q=\pi$ used in Ref.~\citeSM{KSSetal19S}.
In this reference, $P(q,t)$ is defined as $P(q,t) =\sum_j e^{iqR_j} P(j,t)$ with $P(j,t) = \frac{1}{L}\sum_{i} \bra{\Psi(t)} ( \hat{\Delta}^{\dag}_{i+j} \hat{\Delta}_{i} + {\rm c.c.} ) \ket{\Psi(t)}$ and $\hat{\Delta}_j = \hat{c}_{j,\uparrow}\hat{c}_{j,\downarrow}$. 
As the similar results are already reported in Ref.~\citeSM{KSSetal19S}, the external pulse induces an enhancement of $n_d(t)$ and $\braket{\hat{\eta}^2}(t)$. 
The inset of Fig.~\ref{fig:PI-TE_eta-nd} shows the time-dependent $\braket{\hat{\eta}^2}(t)/L$ as the function of $n_d(t)$ at the same time $t$. 
The $\eta$-pairing correlation $\braket{\hat{\eta}^2}(t)/L$ increases monotonically with the time-dependent $n_d(t)$. 
Figure~3 in the main text corresponds to these results with the systematic $L$ dependence study. 

Because of the finite size effect, the optimal photoexcitation frequency weakly depends on the system size.
Figure.~\ref{fig:PI-Ldep_eta-wp} shows the frequency $\omega_p$ dependence of $\braket{\hat{\eta}^2}(t)$ after pumping  with different $L$. 
The peaks of $\braket{\hat{\eta}^2}(t)$ are located at $\omega_p \sim U$ but they have the system size dependence. 
With increasing $L$, the highest peaks of $\braket{\hat{\eta}^2}(t)$ approach to $\omega_p = U$. 
To discuss the finite-size effect systematically, we employ the optimal $\omega_p$, at which the highest peak of $\braket{\hat{\eta}^2}(t)$ is located, for each system size $L$.  

In the main text, we show the time-dependent $\braket{\hat{\eta}^2}(t)/L$ as the function of $n_d(t)$ at a fixed pump strength $A_0$. 
Equivalent results could be obtained by examining $A_0$ dependence of $\braket{\hat{\eta}^2}$ and $n_d$ after pumping.
Figure~\ref{fig:PI-A_eta-nd} shows the time averaged $\braket{\hat{\eta}^2}$ and $n_d$ after pumping. 
The $\eta$-pairing correlation $\braket{\hat{\eta}^2}/L$ increases with $n_d$ as a function of $A_0$. 
Figure~\ref{fig:PI-Ldep_eta-ndSM} summaries $\braket{\hat{\eta}^2}/L$ vs. $n_d$ obtained by varying $A_0$, corresponding to the plot shown in the inset of Fig.~\ref{fig:PI-A_eta-nd} but for different system sizes $L$. 
Note that here we omit the data at $n_d>0.3$ for better visibility. 
As in the case studied in Fig.~3 in the main text, we find that $\braket{\hat{\eta}^2}/L$ in the photoinduced state is much larger than $\braket{\hat{\eta}^2}/L = n_d$ (dashed line) expected for a thermal distribution of the eigenstates [see Eq.~(\ref{eq:eta=nd})] and is enhanced with $L$. 
This suggests the long-ranged $\eta$-pairing correlation in our optically driven system.


\bibliographystyleSM{apsrev4-1}
\bibliographySM{References}

\end{document}